\documentclass[final,twocolumn,12pt]{elsarticle}
\pdfoutput=1
\usepackage{graphicx} \usepackage{amsmath} \usepackage{amsfonts}
\usepackage{amssymb}
\usepackage{comment}
\usepackage{wrapfig}
\usepackage{floatrow}
\usepackage{breqn}
\DeclareFloatFont{cusize}{\tiny}
\def\dO{\partial \Omega}

\def\scriptO{{{\it O}\kern -.42em {\it `}\kern + .20em}}
\def\RR{{{\rm l}\kern - .15em {\rm R} }}
\def\PP{{{\rm l}\kern - .15em {\rm P} }}
\def\L2{{{\sf L}^2}}
\def\H1{{{\sf H}^1}}
\def\PN2{{\PP_{N}-\PP_{N-2}}}
\def\PNN{{\PP_{N}-\PP_{N}}}

\def\complex{{{\rm C} \kern - .53em {\rm l} \kern + .38em}}

\def\a1{{ | \lambda_{\min} |}}

\def\l1{{   \lambda_{\min}  }}
\def\bhu{{\hat   {\bf u}}}

\def\bhu{{\hat   {\bf u}}}
\def\bhn{{\hat   {\bf n}}}

\def\bu0{{\underline {\bf 0}}}

\def\br{{\bf r}}
\def\bu{{\bf u}}

\def\bx{{\bf x}}
\def\bq{{\bf q}}

\def\bhn{{\bf \hat n}}
\def\btu{{\bf \tilde u}}

\def\Oh{{\hat \Omega}}

\def\ue{{\underline e}}

\def\u0{{\underline 0}}

\journal{Computers \& Fluids}

\begin{document}

\begin{frontmatter}



\title{Nonconforming Schwarz-Spectral Element Methods For Incompressible Flow}


\author{Ketan Mittal\corref{cor1}, Som Dutta\corref{cor2}, Paul
Fischer\corref{cor3}} 
\cortext[cor1]{Mechanical Science and Engineering, University of Illinois at
Urbana-Champaign} 
\cortext[cor2]{CSI, City University of New York} 
\cortext[cor3]{Computer Science, University of Illinois at Urbana-Champaign}

\address{}

\begin{abstract}
We present scalable implementations of spectral-element-based Schwarz
overlapping (overset) methods for the incompressible Navier-Stokes (NS)
equations.  Our SEM-based overset grid method is implemented at the level of
the NS equations, which are advanced independently within separate subdomains
using interdomain velocity and pressure boundary-data exchanges at each
timestep or sub-timestep.  Central to this implementation is a general,
robust, and scalable interpolation routine, {\em gslib-findpts}, that rapidly
determines the computational coordinates (processor $p$, element number $e$,
and local coordinates $(r,s,t) \in \Oh := [-1,1]^3$) for any arbitrary  point
$\bx^* =(x^*,y^*,z^*) \in \Omega \subset \RR^3$.  The communication kernels
in {\em gslib} execute with at most $\log P$ complexity for $P$ MPI ranks,
have scaled to $P > 10^6$, and obviate the need for development of any
additional MPI-based code for the Schwarz implementation.  The original
interpolation routine has been extended to account for multiple overlapping
domains. The new implementation discriminates the possessing subdomain by
distance to the domain boundary, such that the interface boundary data is
taken from the inner-most interior points. We present application of this
approach to several heat transfer and fluid dynamic problems, discuss the
computation/communication complexity and accuracy of the approach, and
present performance measurements for $P > 12,000$.  \end{abstract}

\begin{keyword}
Overset \sep High-order \sep Scalability \sep Turbulence \sep Heat-Transfer
\end{keyword}
\end{frontmatter}


\section{Introduction}\label{intro}

High-order spectral element methods (SEMs) are well established as an
effective means for simulation of turbulent flow and heat transfer
in a variety of engineering applications
(e.g., \cite{dfm02,dutta2016,hosseini2016,merzari2017}).
Central to the performance and accuracy of these methods is the use of
hexahedral elements, $\Omega^e$, which are represented by isoparametric mappings
of the reference element $\Oh:=[-1,1]^d$ in $d$ space
dimensions.  With such a configuration, it is possible to express all
operators in a factored matrix-free form that requires only $O(n)$ storage,
where $n=EN^d$ is the number of gridpoints for a mesh comprising $E$ elements
of order $N$.   The fact that the storage scales as $N^d$ and not
$N^{2d}$ is a major advantage of the spectral element method that was put
forth in the pioneering work of Orszag \cite{sao80} and Patera \cite{patera84}
in the early 80s.
The work is also low, $O(N^{d+1})$, and can be cast as dense matrix-matrix
products, which are highly efficient on modern-day architectures \cite{dfm02}.

The efficiency and applicability of the SEM is tied closely to the ability to
generate all-hex meshes for a given computational domain.  While all-tet
meshing is effectively a solved problem, the all-hex case remains challenging
in many configurations.  Here, we explore Schwarz overlapping methods as an
avenue to support nonconforming discretizations in the context of the SEM.
Overlapping grids simplify mesh generation by allowing the user to represent
the solution on a complex domain as grid functions on relatively simpler
overlapping regions (also known as \emph{overset} or \emph{chimera} grids).
These simpler overlapping regions allow grid generation where local mesh
topologies are otherwise incompatible, which is a feature of particular
importance for complex 3D domains and in the simulation of flows in
time-dependent domains subject to extreme mesh deformation. Overlapping grids
also enable use of discretization of varying resolution in each domain based on
the physics of the problem in that region.

\begin{figure}[t]
\begin{center}
$\begin{array}{ccc}
\includegraphics[height=15mm,width=24mm]{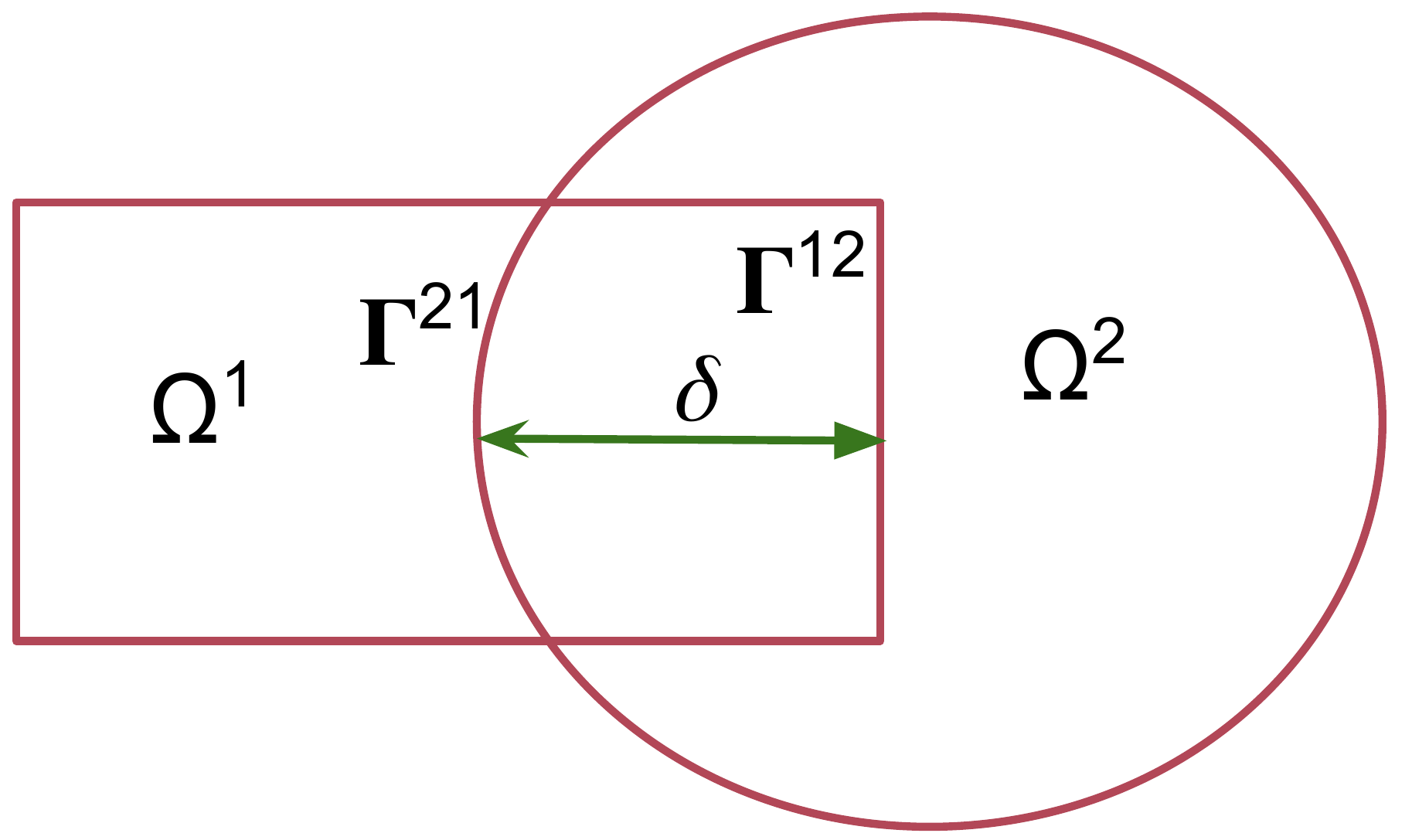} & \hspace{2mm} 
\includegraphics[height=15mm,width=15mm]{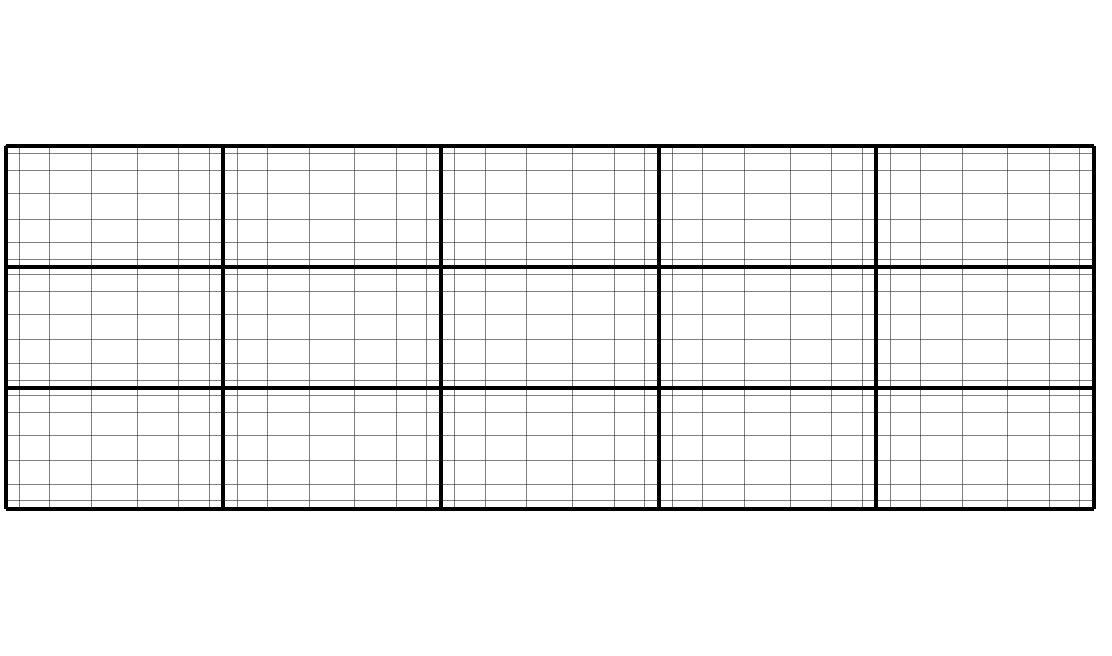} &\hspace{-2mm}
\includegraphics[height=15mm]{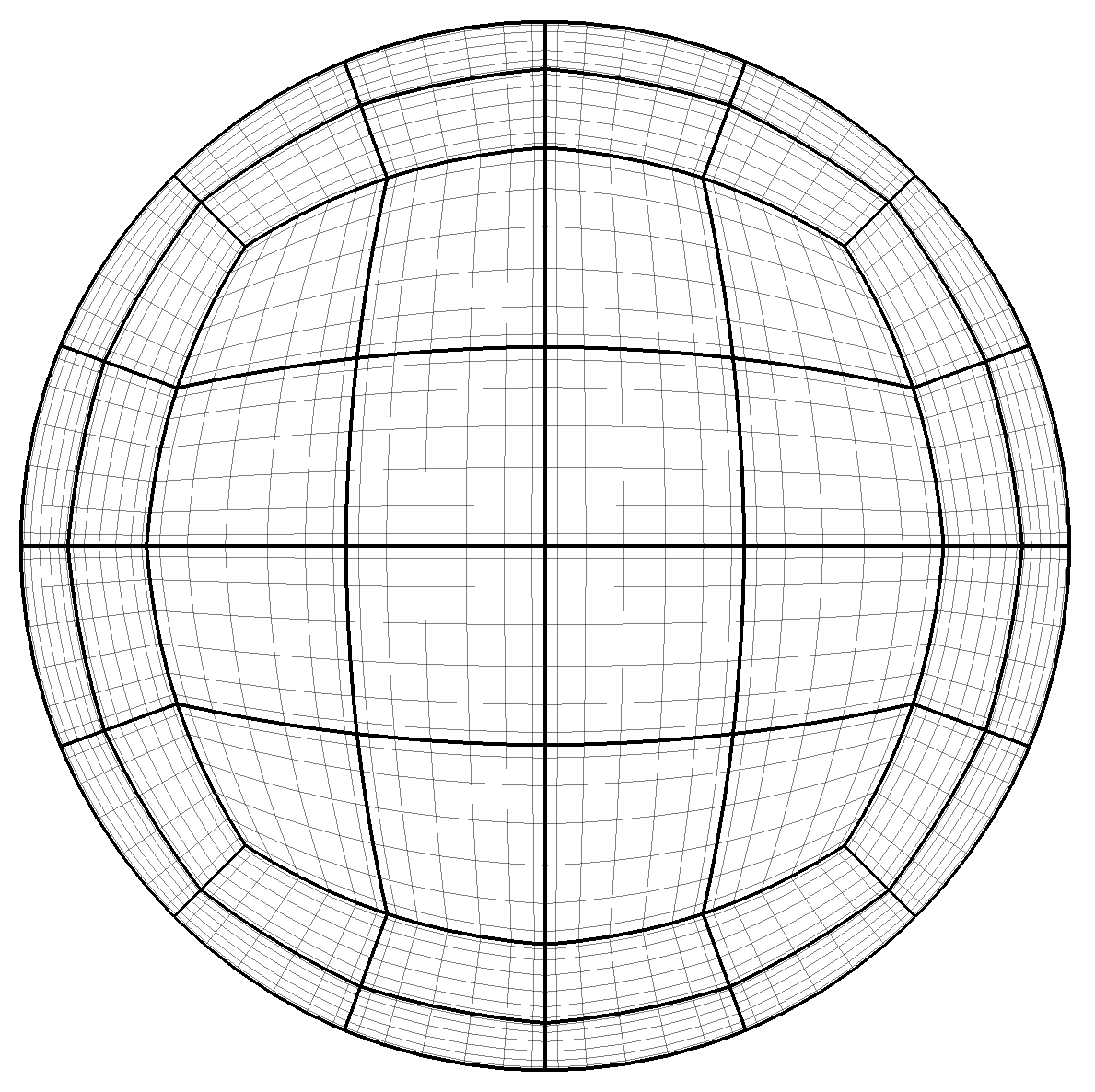} \\
\end{array}$
\end{center}
\vspace{-7mm}
\caption{Overlapping rectangular and circular subdomains with corresponding
spectral element meshes and Gauss-Lobatto-Legendre grids for $N=7$.
$\Gamma^{ij}$ denotes the segment of the subdomain boundary $\dO^i$ that is
interior to $\Omega^j.$ \vspace*{-.18in}} 
\label{fig:schwarz}
\end{figure}

Overlapping grids introduce a set of challenges that are not posed by a single
conforming grid \cite{Rogersbestpractices}. Using multiple meshes requires
boundary condition for each \emph{interface/artificial} boundary that is
interior to the other domain ($\Gamma^{12}$ and $\Gamma^{21}$ in Fig.
\ref{fig:schwarz}). Since these surfaces are not actual domain boundaries,
boundary condition data must be interpolated from a donor element in the
overlapping mesh, which must be identified at the beginning of a calculation.
For moving or deforming meshes, donor elements must be re-identified after each
timestep. Additionally, if multiple overlapping meshes share a target donor
point, it is important to pick the donor mesh that will provide the most
accurate solution. For production-level parallel computing applications, the
identification of donor element, interpolation of data, and communication must
be robust and scalable. Additionally, differences in resolution of the
overlapping meshes can impact global mass-flux balances, leading to a violation
of the divergence-free constraint with potentially stability consequences for
incompressible flow calculations. These issues must be addressed in order to
enable scalable and accurate turbulent flow and heat transfer calculations.

Since their introduction in 1983 \cite{steger1983}, several overset methods
have been developed for a variety of problems ranging from Maxwell's Equations
\cite{angel2018} to fluid dynamics \cite{chandar2018comparative} to particle
tracking \cite{koblitz2017}.  Overset-grid methods are also included in
commercial and research software packages such as Star-CCM \cite{cd2012v7},
Overflow \cite{nicholsoverflowmanual} and Overture \cite{overturemanual}.  Most
of these implementations, however, are at most fourth-order accurate in space
\cite{angel2018,koblitz2017,coder2017contributions,henshaw1994,brazell2016overset,crabill2016}.
Sixth-order finite-difference based methods have been presented in
\cite{aarnes2018high,nicholsoverflowmanual}, while sixth-order finite volume
based schemes are available in elsA \cite{cambier2013onera}. A spectral element
based Schwarz method presented by Merrill et. al.  \cite{merrill2016} has
demonstrated exponential convergence.

Here, we extend the work of Merrill et. al. \cite{merrill2016} to develop a
robust, accurate, and scalable framework for overlapping grids based on the
spectral element method.  We start with a brief description of the SEM and
existing framework which we build upon (Section \ref{sec: sembackground}).  We
discuss our approach to fast, parallel, high-order interpolation of boundary
data and a mass-balance correction that is critical for incompressible flow.
Finally we present some applications in Section \ref{sec: apps} which
demonstrate the scalability and accuracy of our overlapping grid framework.

\section{SEM-Schwarz for Navier-Stokes} \label{sec: sembackground}

The spectral element method (SEM) was introduced by Patera for the solution of
the incompressible Navier-Stokes equations (NSE) in \cite{patera84}.
The geometry and the solution are represented using the $N$th-order
tensor-product polynomials on isoparametrically mapped elements.  Variational
projection operators are used to discretize the associated partial differential
equations.  While technically a $Q_N$ finite element method (FEM), the SEM's
strict matrix-free tensor-product formulation leads to fast implementations
that are qualitatively different than classic FE methods.  SE storage is only
$O(n)$ and work scales as $O(EN^{d+1})=O(nN)$, where $n=EN^d$ is the number of
points for an $E$-element discretization of order $N$ in $\RR^d$.
(Standard $p$-type FE formulations have work and storage complexities of
$O(nN^d)$, which is prohibitive for $N>3$.)
All $O(N^{d+1})$ work terms in the SEM can be cast as fast matrix-matrix products. 
A central feature of the SEM is to use nodal bases on the Gauss-Lobatto-Legendre
(GLL) points, which lead to an efficient and accurate (because
of the high order) diagonal mass matrix.  Overintegration (dealiasing) is
applied only to the advection terms to ensure stability \cite{malm13}.

For the unsteady NSE, we use semi-implicit BDF$k$/EXT$k$
timestepping in which the time derivative is approximated by a $k$th-order
backward difference formula (BDF$k$), the nonlinear terms (and any other
forcing) are treated with $k$th-order extrapolation (EXT$k$), and the viscous,
pressure, and divergence terms are treated implicitly.  This
approach leads to a linear unsteady Stokes problem to be solved at each
timestep, which is split into independent viscous and pressure (Poisson)
updates, with the pressure-Poisson problem being the least well conditioned
(and most expensive) substep in the time advancement.  We support two spatial
discretizations for the Stokes problem: the $\PN2$ formulation
with a continuous $Q_N$ velocity space and a discontinuous $Q_{N-2}$
pressure space; and the $\PNN$ approach having equal order continuous
velocity and pressure.  Full details are provided in \cite{dfm02,fischer17}.

The Schwarz-SEM formulation of the NSE brings forth several considerations.
Like all Schwarz methods, the basic idea is to interpolate data on subdomain
boundaries that are interior to $\Omega$ from donor subdomains and
to then solve the subdomain problems independently (in parallel) on different
processor subsets.  In principle, the NSE require only velocity boundary
conditions.  The first challenge is to produce that boundary data in an
accurate and stable way.  Spatial accuracy comes from using high-order
(spectral) interpolation, as described in the next section.  


For temporal accuracy, there are several new concerns. As a cost-saving
measure, one can simply use lagged donor-velocity values (corresponding to
piecewise-constant extrapolation) for the subdomain interface updates.  While
only first-order accurate, this scheme is stable without the need for
subiterations. To realize higher-order accuracy (up to $k$), one can
extrapolate the boundary data in time.  Generally, this extrapolation must be
coupled with a predictor-corrector iteration for the unsteady-Stokes substep.
(The nonlinear term is already accurately treated by extrapolation and is
stable provided one adheres to standard CFL stability limits.)  Typically three
to five subiterations ($\kappa_{iter}$) are needed per timestep to ensure
stability \cite{peet2012} for $m$th-order extrapolation of the interface
boundary data, when $m>1$.  Using this approach, Merrill {\em et al.}
\cite{merrill2016} demonstrated exponential convergence in space and up to
third-order accuracy in time for Schwarz-SEM flow applications.

From a practical standpoint, our SEM domain decomposition approach is enabled
by using separate MPI communicators for each overlapping mesh, $\Omega^s$,
which allows all of the existing solver technology (100K lines of code) to
operate with minimal change.  The union of these separate {\em session} (i.e.,
subdomain)
communicators is {\tt MPI\_COMM\_WORLD}, which is invoked for the subdomain data
exchanges and for occasional collectives such as partition-of-unity-weighted
integrals over $\Omega=\bigcup \Omega^s$.  The data exchange and donor point-set
identification is significantly streamlined through the availability of a
robust interpolation library, \emph{gslib-findpts}, which obviates the need
for direct development of any new MPI code, as discussed next.

\subsection{Interpolation} \label{sec: findpts}
The centerpiece of our multidomain SEM-based nonconforming Schwarz code is the
fast and robust interpolation utility {\em findpts}.  This utility grew out of
a need to support data interrogation and Lagrangian particle tracking on
$P=10^4$--$10^6$ processors. High fidelity interpolation for highly curved
elements, like the ones supported by SEM, is quite challenging. Thus, {\em
findpts} was designed with the principles of robustness (i.e., it should never
fail) and speed (i.e., it should be fast).

\emph{findpts} is part of \emph{gslib}, a lightweight \emph{C} communication
package that readily links with any \emph{Fortran}, {\em C}, or \emph{C++} code.  
\emph{findpts} provides two key capabilities. 
First, it determines computational coordinates $\bq^*=(e^*,p^*,r^*,s^*,t^*)$
(element $e$, processor $p$ and reference-space coordinates $\br = (r,s,t)$)
for any given point $\bx^* = (x^*,y^*,z^*) \in \RR^3$.  
Second, \emph{findpts\_eval} interpolates any given scalar field for a 
given set of computational coordinates (determined from \emph{findpts}).
Because interpolation is nonlocal (a point $\bx^*$ may originate from any
processor), \emph{findpts} and \emph{findpts\_eval} require interprocessor
coordination and must be called by all processors in a given communicator.
For efficiency reasons, interpolation calls are typically batched with all
queries posted in a single call, but the work can become serialized if all
target points are ultimately owned by a single processor.  All communication
overhead is at most $\log P$ by virtue of {\em gslib}'s generalized and
scalable\footnote{We note that {\tt mpi\_alltoall} is {\em not} scalable 
because its interface requires arguments of length $P$ on each
of $P$ processors, which is prohibitive when $P>10^6$.}
all-to-all utility, {\em gs\_crystal}, which is based on the {\em crystal
router} algorithm of \cite{fox88}.

To find a given point $\bq^*(\bx^*)$, {\em findpts} first uses a hash table to
identify processors that could potentially own the target point $\bx^*$.
A call to {\em gs\_crystal} exchanges copies of the $\bx^*$ entries between
sources and potential destinations.   Once there, elementwise bounding boxes
further discriminate against entries passing the hash test.  At that point, a
trust-region based Newton optimization is used to minimize
$||\bx^{e}(\br)-\bx^*||^2$ and determine the computational coordinates for that
point.  The initial guess in the minimization is taken to be the closest point
on the mapped GLL mesh for element $e$.  In addition to returning the
computational coordinates of a point, \emph{findpts} also indicates whether a
point was found inside an element, on a border, or was not found within the
mesh. Details of the \emph{findpts} algorithm are given in
\cite{findpts}. 

In the context of our multidomain Schwarz solver, {\em findpts} is called at
the level of {\tt MPI\_COMM\_WORLD}.  One simply posts all meshes to {\em
findpts\_setup} along with a pair of discriminators.  The first discriminator
is an integer field which, at the setup and execute phases, is equal to the
subdomain number (the {\em session id}).  In the {\em findpts} setup phase,
each element in $\Omega$ is associated with a single subdomain $\Omega^j$,
and the subdomain number $j$ is passed in as the discriminator.  During the 
{\em findpts} execute phase, one needs boundary values for interface points on
$\Omega^j$, but does not want these values to be derived from elements in
$\Omega^j$.  All interface points associated with $\dO^j$ are tagged with the
$j$ discriminator and {\em findpts} will only search elements in 
$\Omega \backslash \Omega^j$.  

\begin{figure}[h] \begin{center} 
\includegraphics[height=40mm]{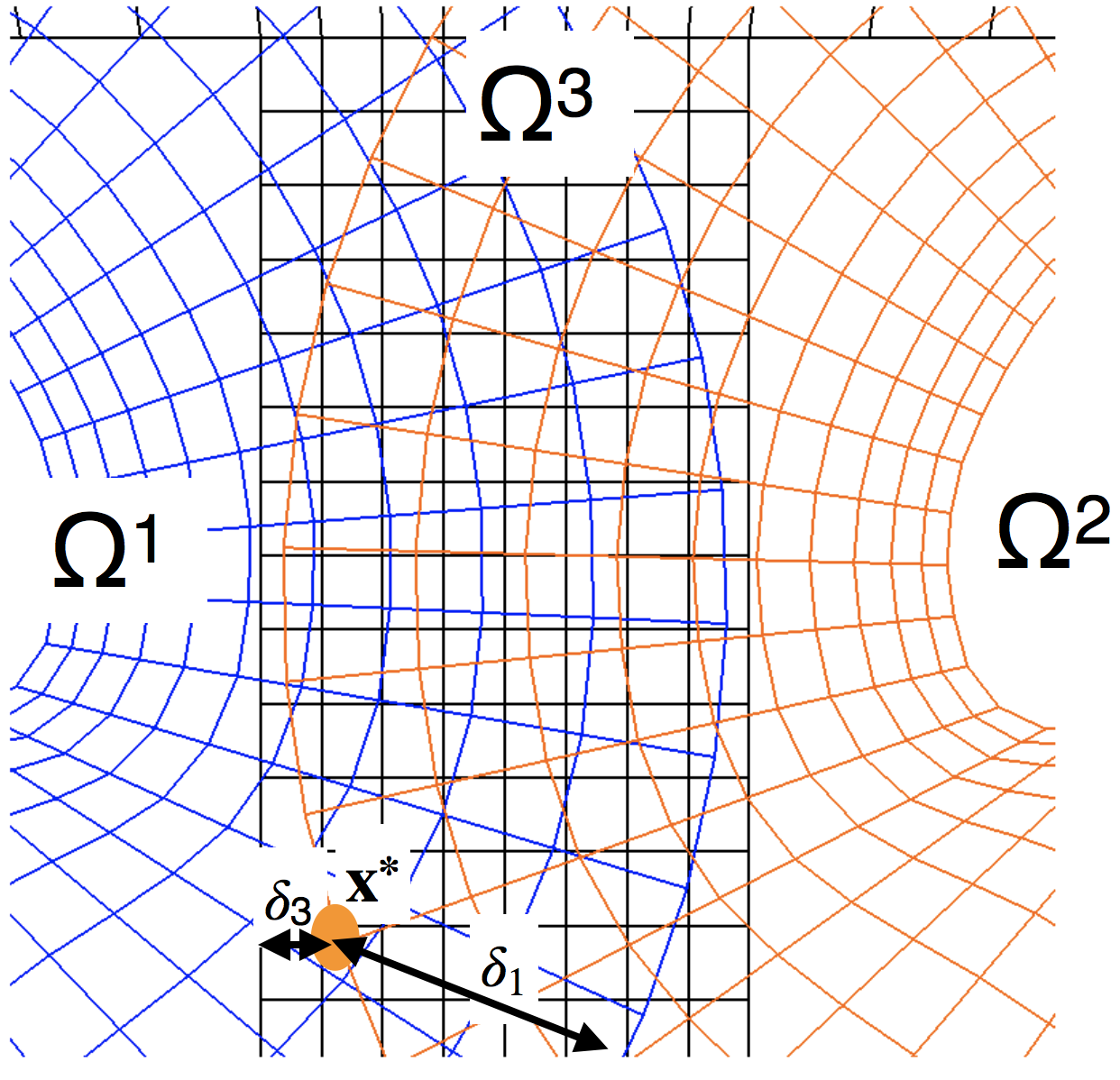}     
\vspace{-7mm}
\caption{\emph{findpts} considers distance from the interface boundaries
($\delta^{1}$ in $\Omega^1$ and $\delta^{3}$ in $\Omega^3$) when determining
the best donor element for $\bx^* \in \dO^2.$} 
\end{center} \label{fig:neknekdom} \end{figure}

In the case of multiply-overlapped domains, it is still possible to have more
than one subdomain claim ownership of a given boundary point $\bx^*$.  To
resolve such conflicts, we associate with each subdomain $\Omega^j$ a local
distance function, $\delta^j(\bx)$, which indicates the minimum distance from any
point $\bx \in \Omega^j$ to $\dO^j$.  The ownership of any boundary point
$\bx^* \in \dO^j$ between two or more domains $\Omega^k$, $k\neq j$, is taken
to be the domain that maximizes $\delta^k(\bx^*)$.  This choice is motivated by
the standard Schwarz arguments, which imply that errors decay as one moves {\em
away} from the interface, in accordance with decay of the associated Green's
functions.  We illustrate this situation in Fig. \ref{fig:neknekdom} where
$\bx^* \in \dO^2$ belongs to $\Omega^1$ and $\Omega^3$.  In this case, 
interpolated values (from {\em findpts\_eval}) will come from $\Omega^1$
because $\bx^*$ is ``more interior'' to $\Omega^1$ than $\Omega^3$.

\subsection{Mass-Balance}
Since pressure and divergence-free constraint are tightly coupled in
incompressible flow, even small errors at interface boundaries (due to
interpolation or use of overlapping meshes of disparate scales) can lead to
mass-imbalance in the system, resulting in erroneous and unsmooth pressure
contours.  For a given subdomain $\Omega^j$, the mass conservation statement
for incompressible flow is simply
\begin{eqnarray} \label{eq:consv1}
\int_{\dO^j} \bu \cdot \bhn &=&0,
\end{eqnarray}
where $\bhn$ represents the outward pointing unit normal vector on $\dO^j$.
Our goal is to find a nearby correction to the interpolated surface data
that satisfies (\ref{eq:consv1}).

Let $\dO_D$ denote the subset of the domain boundary $\dO$ corresponding to
Dirichlet velocity conditions and $\dO_N$ be the Neumann (outflow) subset.  If
$\dO_N \cap \dO^j=0$, then there is a potential to fail to satisfy
(\ref{eq:consv1}) because the interpolated fluxes on $\dO^j$ may not integrate
to zero.  Let $\bhu$ denote the tentative velocity field defined on
$\dO^j$ through prescribed data on $\dO_d := \dO^j \cap \dO_D$ and
interpolation on $\dO_i := \dO^j \backslash \dO_d$.
Let $\bu = \bhu + \btu$ be the flux-corrected boundary data on $\dO^j$
and $\btu$ be the correction required to satisfy (\ref{eq:consv1}).
One can readily show that the choice
\begin{eqnarray} \label{eq:corr}
\left.  \btu \right|^{}_{\dO_i}
&=& \left.  \gamma \bhn \right|^{}_{\dO_i}
\end{eqnarray}
is the $L^2$ minimizer of possible trace-space corrections
that allow (\ref{eq:consv1}) to be satisfied, provided that
\begin{eqnarray} \label{eq:consv2} \nonumber
\gamma 
&=&
- \frac{ \int_{\dO_d} \bhu \cdot \bhn \, dA   +
         \int_{\dO_i} \bhu \cdot \bhn \, dA }
       { \int_{\dO_i} \bhn \cdot \bhn \, dA } \\[1ex]
&=&
- \frac{ \int_{\dO^j} \bhu \cdot \bhn \, dA }
       { \int_{\dO_i} \bhn \cdot \bhn \, dA }.
\end{eqnarray}
The denominator of (\ref{eq:consv2}) of course equates to the
surface area of the interface boundary on $\Omega^j$.
The correction (\ref{eq:corr}) is imposed every time boundary data is
interpolated between overlapping domains during Schwarz iterations.

\section{Results \& Applications}  \label{sec: apps}
We now present several applications that demonstrate the
performance and accuracy of the Schwarz-SEM implementation.

\subsection{Parallel Performance}
A principal concern for the performance of Schwarz methods is the overhead
associated with interpolation, especially for high-order methods, where 
interpolation costs scale as $O(N^3)$ per interrogation point in $\RR^3$.
Our first examples address this question.

Figure 3 shows a test domain consisting of a single
spectral element ($N=15$) in $\RR^3$. The spiral configuration leads to many local
mimima in the Newton functional, but the use of nearest GLL points as initial
guesses generally avoids being trapped in false minima. On a 2.3 GHz Macbook
Pro, {\em findpts} requires $\approx$ 0.08 seconds to find $\bq$ for 1000
randomly distributed points on $\hat{\Omega}$ to a tolerance of $10^{-14}$.

\begin{figure}[h] \begin{center}
\includegraphics[height=45mm]{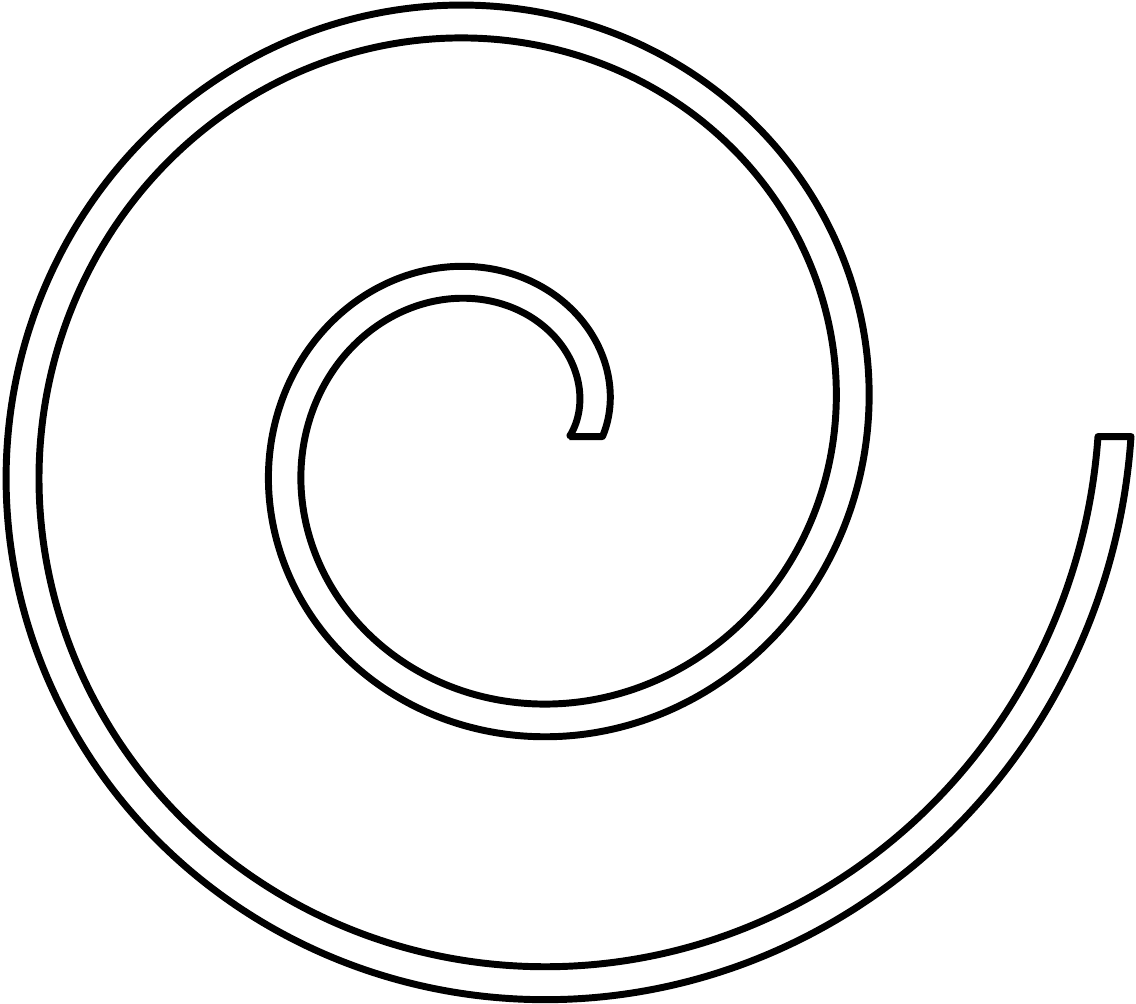}
\vspace{-7mm}
\caption{2D slice of a 3D spiral at $N=15$. \emph{findpts} robustly finds 1000
random points in $\approx$ 0.08 sec.} \end{center} \label{fig:spiral}
\end{figure}

\begin{figure}[h]
\begin{center}
$\begin{array}{cc}
\includegraphics[height=35mm,width=35mm]{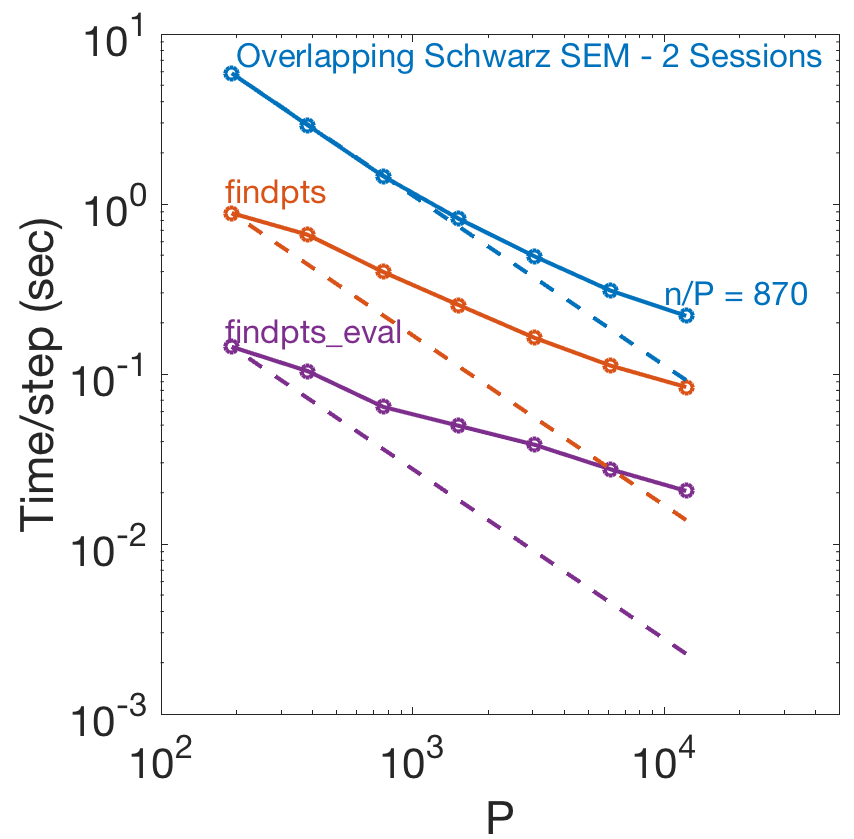} &  
\includegraphics[height=35mm,width=35mm]{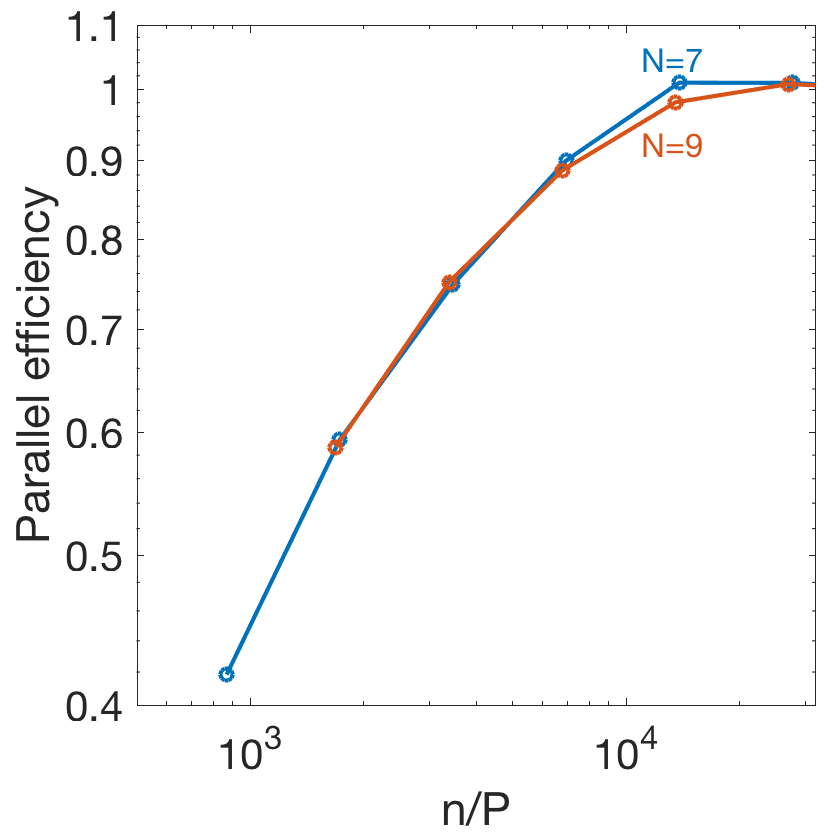} \\
\end{array}$
\end{center}
\vspace{-7mm}
\caption{(left) Strong scaling plot at $N=7$, and (right) parallel
efficiency of the method at $N=7$ and $N=9$.}
\label{fig:scaling}
\end{figure}

\begin{figure}[h]
\begin{center}
$\begin{array}{cc}
\includegraphics[height=35mm,width=35mm]{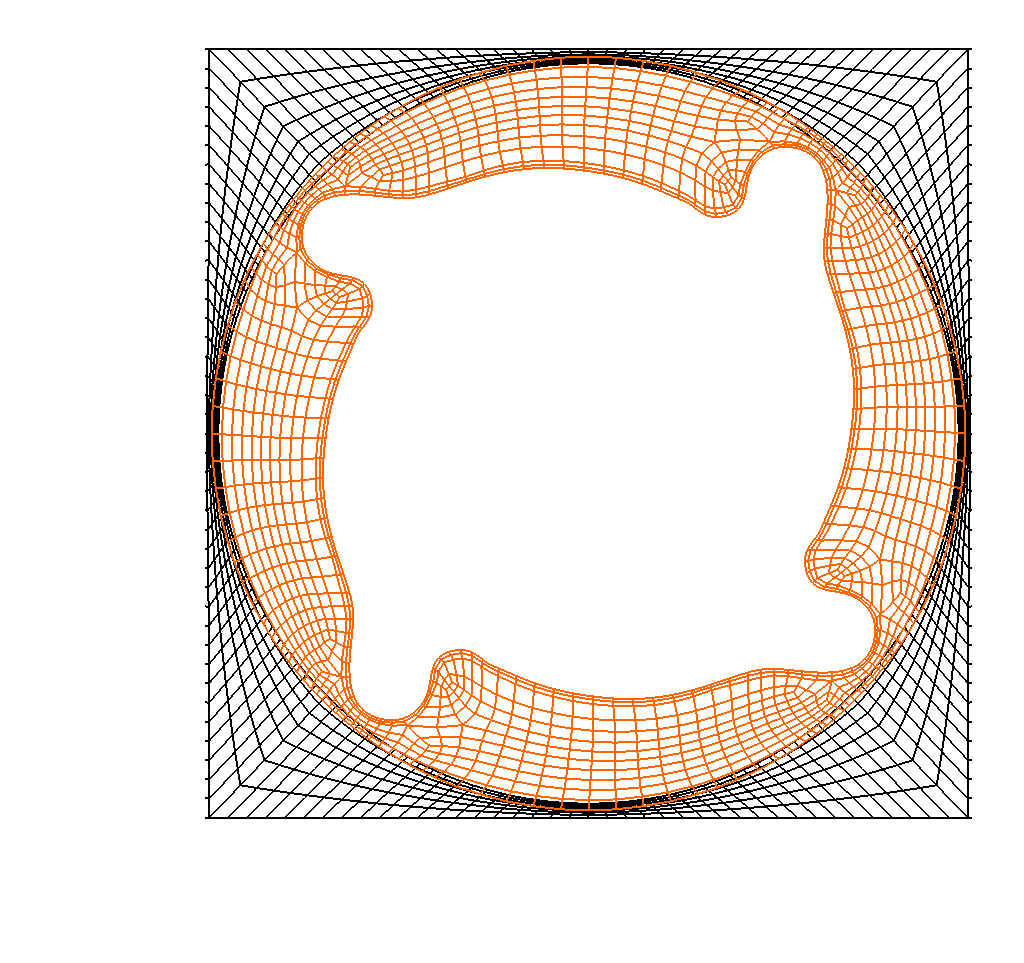} &  
\includegraphics[height=35mm,width=35mm]{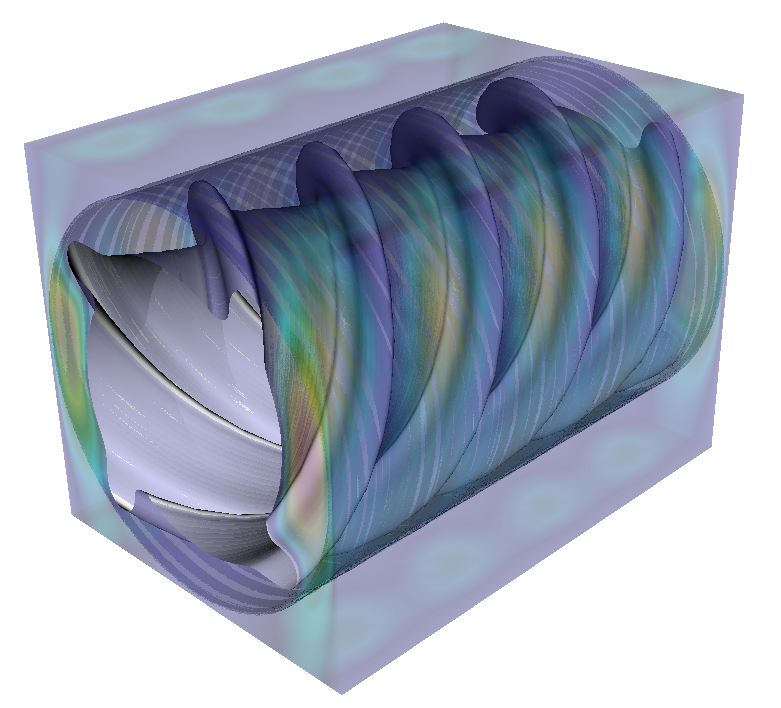} \\
\end{array}$
\end{center}
\vspace{-7mm}
\caption{ (left) Overlapping 2D spectral element meshes for the
notched-cylinder geometry, and (right) velocity magnitude contours highlighting 
the twisted cylinder.}
\label{fig:knotch}
\end{figure}

In addition to the {\em findpts} cost, we must be concerned with the overhead
of the repeated {\em findpts\_eval} calls, which are invoked once per
subiteration for each Schwarz update.  In a parallel setting, interpolation
comes with the additional overhead of communication, which is handled
automatically in $\log P$ time by {\em findpts\_eval}.  Figure
\ref{fig:scaling} shows a strong-scale plot of time versus number of processors
$P$ for a calculation with $E$=20,000 spectral elements at $N=7$ ($n$=6.8
million grid points) on Cetus, an IBM Blue Gene/Q at the Argonne Leadership
Computing Facility. For this geometry, shown in Fig.  \ref{fig:knotch}, there
are 2000 elements at the interface-boundary with a total of 128,000
interface points that require interpolation at each Schwarz iteration.
Overlapping domains simplify mesh generation for this problem by allowing an
inner mesh (twisting in the spanwise direction) to overlap with an extruded
outer mesh.

Figure \ref{fig:scaling} shows scaling for overall time per step and time spent
in \emph{findpts} and \emph{findpts\_eval}. Due to the inherent load imbalance
in overlapping grids, owing to the fact that all interface elements might not
be located on separate processors, the scaling for \emph{findpts} and
\emph{findpts\_eval} is not ideal. However, \emph{findpts} takes only 10\% and
\emph{findpts\_eval} takes 1\% of time compared to the total time to solution
per timestep, and as a result the scaling of the overall method is maintained.
The parallel efficiency of the calculation is more than $90\%$ until the number
of MPI ranks exceeds the number of interface elements. The parallel efficiency
drops to $60\%$ at $n/P = 1736$, which is in accord with performance models for
the (monodomain) SEM \cite{fischer2015scaling}.

\subsection{Exact Solution for Decaying Vortices}
\begin{figure}[t]
\begin{center}
$\begin{array}{cc}
\includegraphics[height=35mm,width=35mm]{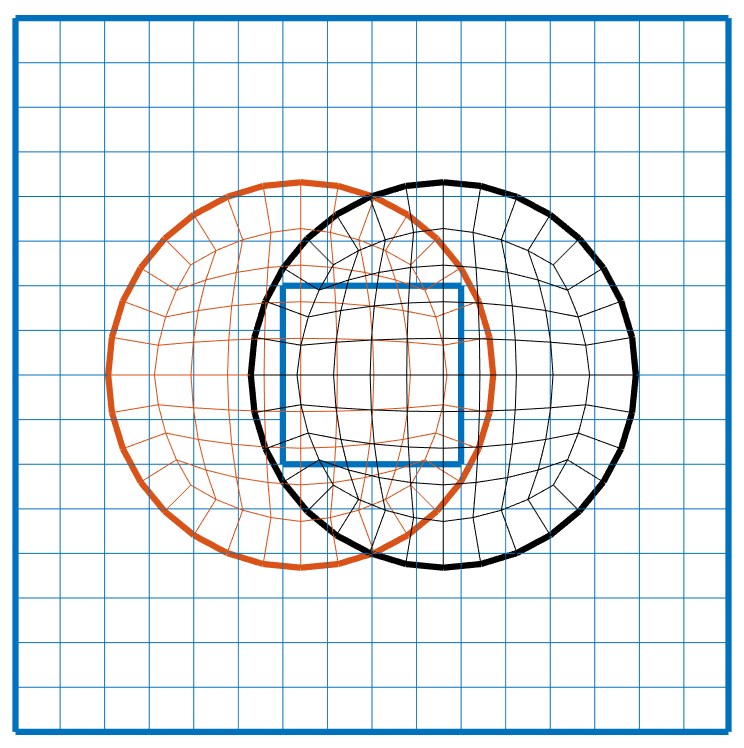} & 
\includegraphics[height=35mm,width=35mm]{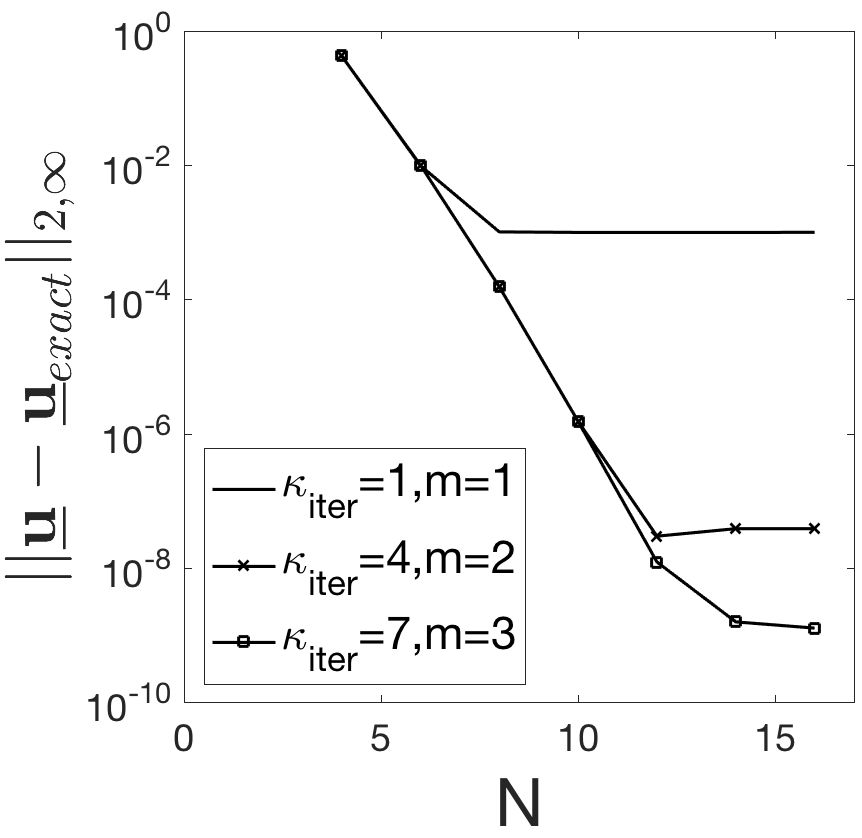} \\
\end{array}$
\end{center}
\vspace{-7mm}
\caption{Walsh Eddy Problem: 
(left) three overlapping spectral element meshes and 
(right) spectral convergence for error in velocity.}
\label{fig:eddy_nnn} \end{figure}

Our next example demonstrates that the Schwarz implementation
preserves the exponential convergence expected of the SEM.
Walsh derived a family of exact eigenfunctions for the Stokes and Navier-Stokes
equations based on a generalization of Taylor-Green vortices in the periodic
domain $\Omega = [0,2\pi]^2$.  For all integer pairs ($m, n$) satisfying
$\lambda = -(m^2+n^2)$, families of eigenfunctions can be formed by defining
streamfunctions $\psi$ that are linear combinations of
$\cos(mx) \cos(ny),$  
$\sin(mx) \cos(ny),$
$\cos(mx) \sin(ny),$ and
$\sin(mx) \sin(ny).$Taking as an initial condition the eigenfunction $\hat{\bu} = (-\psi_y,\psi_x)$, a 
solution to the NSE is $\bu = e^{\nu \lambda t}
\hat{\bu}(\bx)$.  The solution is stable only for modest Reynolds numbers.
Interesting long-time solutions can be realized, however, by adding a
relatively high-speed mean flow to the eigenfunction.  We demonstrate the
multidomain capability using the three meshes illustrated in Fig.
\ref{fig:eddy_nnn} (left); a periodic background mesh has a square hole in the
center while a pair of circular meshes cover the hole.  

Exponential convergence of the velocity error with respect to $N$ is
demonstrated in Fig.  \ref{fig:eddy_nnn} (right).  (Here, the norm is the
pointwise maximum of the 2-norm of the vector field, i.e., $||{\bf
\ue}||_{2,\infty} := \max_{i} ||{\bf e}_i||_2$.)  For extrapolation
orders $m=1$, 2, and 3, $\kappa_{iter}$ was set to 1, 4 and 7 to ensure
stability.

\subsection{Vortex Breakdown}
\begin{figure}[h]
\begin{center}
$\begin{array}{cc}
\includegraphics[height=30mm]{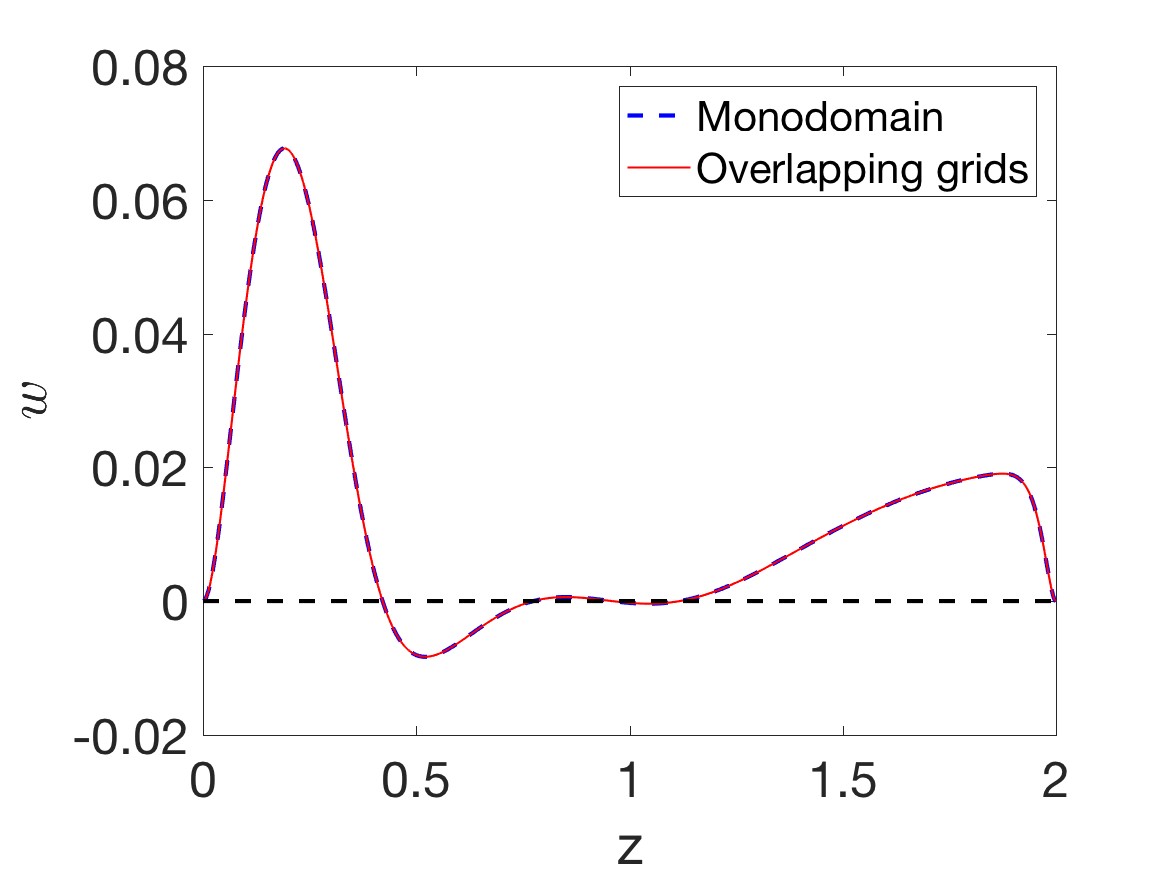} & 
\includegraphics[height=30mm]{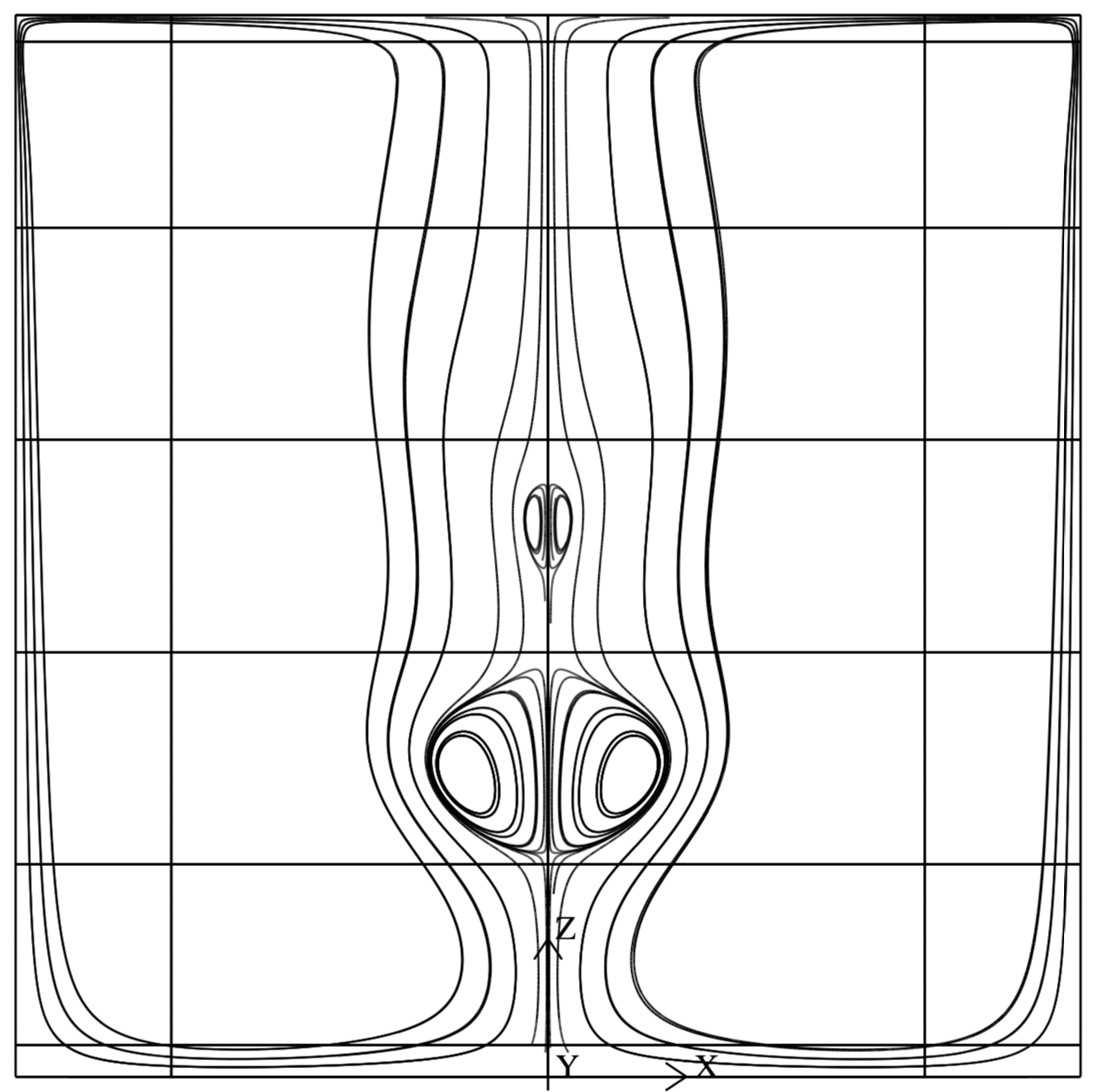} \\
\end{array}$
\end{center}
\vspace{-7mm}
\caption{ Vortex breakdown problem: (left) $w$ along cylinder centerline
and (right) in-plane streamlines.} \label{fig:vbnnvsn}
\end{figure}

Escudier \cite{escudier1984} studied vortex breakdown in a cylindrical
container with a lid rotating at angular velocity $\Omega$. He considered
cylinders for various different $H/R$ i.e. height to radius ratio, at
different $Re = \frac{\Omega^2 R}{\nu}$.  Sotiropoulos \& Ventikos
\cite{sotiropoulos1998} did a computational study on this experiment comparing
the structure and location of the bubbles that form as a result of vortex
breakdown.  Here, we use overlapping grids for $Re=1854$ and $H/R = 2$ case to
compare our results against \cite{escudier1984}, \cite{sotiropoulos1998}, and a
monodomain SEM based solution.  The monodomain mesh has 140 elements at $N=9$.
The overlapping mesh was generated by cutting the monodomain across the
cylinder axis and extruding the two halves.  The calculations were run with
$m=1$ and $\kappa_{iter}=1$ (no subiteration) for 2000 convective time-units to
reach steady state.
Figure \ref{fig:vbnnvsn} compares the axial velocity along the centerline for
monodomain and overlapping grids based solution, and Table \ref{table: vbnnvsn}
compares these solution with Escudier and Sotiropoulos.   At this resolution,
the Schwarz-SEM and SEM results agree to within $<$ 1 percent and
to within 1 to 2 percent of the results of \cite{sotiropoulos1998}.

\begin{table}[h]
\begin{center}
 \begin{tabular}{|c | c | c | c | c|} 
 \hline
  & $z_1$ &  $z_2$ &  $z_3$ &  $z_4$ \\
 \hline\hline
 Overlapping & 0.4222  & 0.7752  & 0.9596 & 1.1186 \\
 \hline
 Monodomain & 0.4224    & 0.7748   & 0.9609 &  1.1191 \\ 
 \hline
 Escudier & 0.42 & 0.74 & 1.04 & 1.18 \\
 \hline
 Sotiropoulos & 0.42 & 0.772 & 0.928 & 1.09 \\
 \hline
\end{tabular}
\vspace{-7mm} \caption{Zero-crossings ($z$) for vertical velocity along cylinder
centerline for the vortex breakdown.}
\label{table: vbnnvsn}
\end{center}
\end{table}
 
\subsection{Turbulent Channel Flow}
Turbulent boundary layers are one of the applications where
overlapping grids offer the potential for significant savings.  These flows
feature fine scale structures near the wall with relatively larger scales in
the far-field.  As a first step to addressing this class of problems, we
validate our Schwarz-SEM scheme for turbulent channel flow, for which abundant
data is available in literature.  In particular, we compare mono- and
multidomain SEM results at Reynolds number
$Re_{\tau}=\frac{u_{\tau}h}{\nu}=180$ with direct numerical simulation (DNS)
results of Moser et. al.  \cite{moser1999}, who used 2.1 million grid points,
and with the DNS of Vreman \& Kuerten \cite{vreman2014}, which used 14.2
million gridpoints. The Reynolds number is based on the friction velocity
$u_{\tau}$ at the wall, channel half-height $h$ and the fluid kinematic
viscosity $\nu$, with $u_{\tau} = \sqrt{\frac{\tau_w}{\rho}}$ determined using
the wall shear stress $\tau_w$ and the fluid density $\rho$.

Table \ref{table: dns180par} lists the key parameters for the four different
calculations. Following \cite{moser1999} and \cite{vreman2014}, the streamwise
and spanwise lengths of the channel were $4\pi h$ and
$4\pi h/3$, respectively.  Statistics for all SEM results were collected over 50
convective time units.

\begin{table}[h]
\begin{center}
 \begin{tabular}{|p{13mm} |c|c| p{18mm}|}
 \hline
  & $Re_{\tau}$ & $\Omega_y$ & Grid-size \\  
 \hline\hline
 Monodomain    & 179.9 & $[-1,1]$ &  $ 18 \times 18 \times 18 \times N^3$\\ 
\hline
 Overlapping    & 179.9 & $[-1,-0.88]$ &  $19 \times 4 \times 19 \times N^3$\\
                &       & $[-0.76,1]$  &  $18 \times 15  \times 18 \times N^3$\\
\hline
 Moser                  & 178.1 & $[-1,1]$    & $ 128 \times 129 \times 128$\\
 \hline
 Vreman                 & 180   & $[-1,1]$    & $ 384 \times 193 \times 192$\\
 \hline
\end{tabular}
\vspace{-7mm}
\caption{Parameters for channel flow calculations.}
\label{table: dns180par}
\end{center}
\end{table}

\begin{figure}[h]
\begin{center}
$\begin{array}{cc}
\includegraphics[height=35mm,width=35mm]{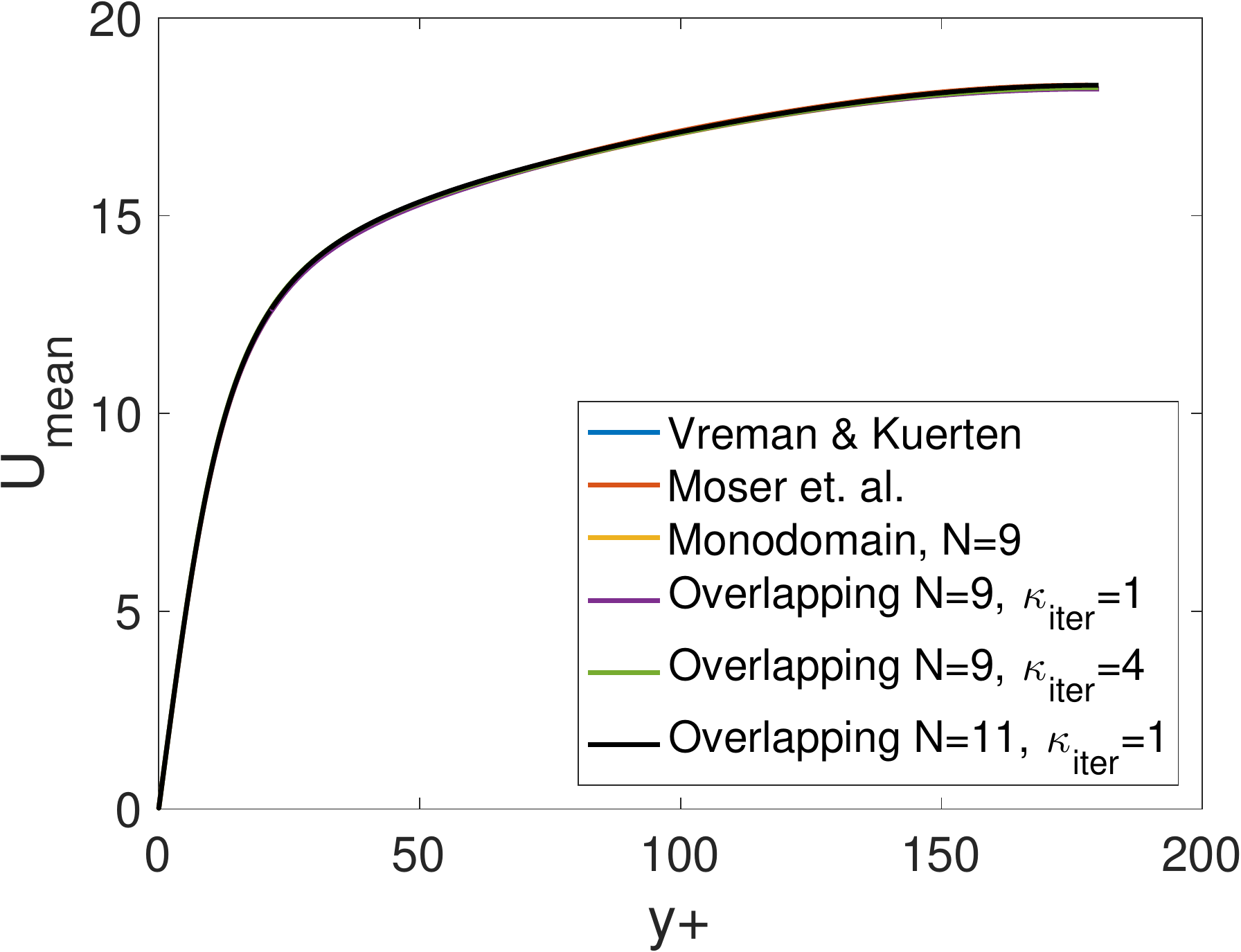} & 
\includegraphics[height=35mm,width=35mm]{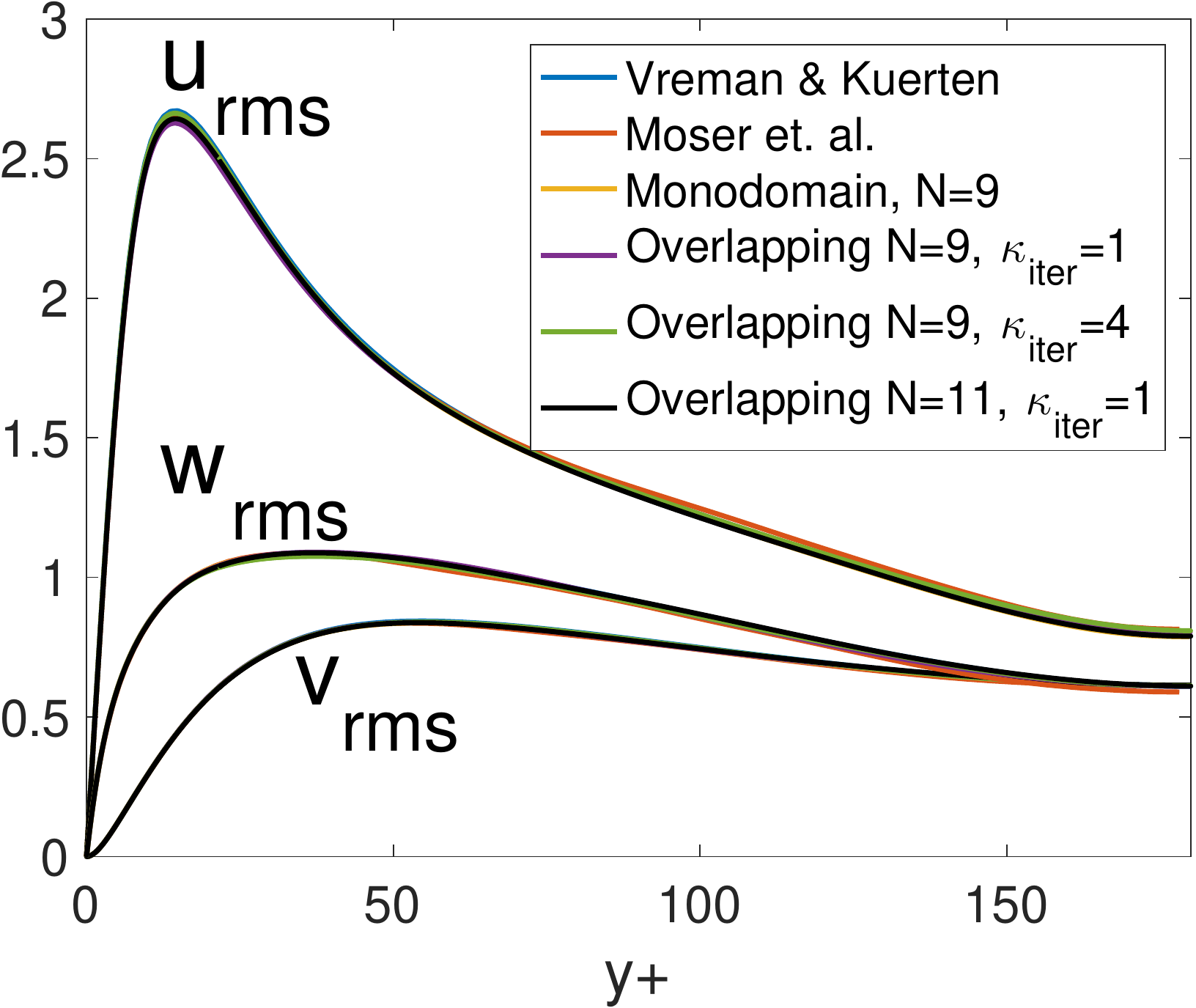} \\
\end{array}$
\end{center}
\vspace{-7mm}
\caption{ Comparison of monodomain, overlapping grid, and Moser et. al. with
Vreman \& Kuerten for $U_{mean}$ (left) and $u_{rms}$, $v_{rms}$ and $w_{rms}$ at
$N=7$ and $N=9$.} \label{fig: chandns}
\end{figure}

Figure \ref{fig: chandns} shows the mean streamwise velocity ($U_{mean}$) and
fluctuations ($u_{rms}$, $v_{rms}$ and $w_{rms}$) versus 
$y^+= \frac{u_{\tau} y}{\nu}$ for each case, where $y$ is the
distance from the nearest wall. For the Schwarz case, several combinations of
resolution ($N$) and subiteration counts ($\kappa_{\mbox{iter}}$) were
considered.  We quantify the relative error for each quantity by computing the
norm of the relative percent difference from the results of Vreman \& Kuerten.
Specifically, define 
\begin{eqnarray}
\epsilon_{\psi}(y) 
   &:=& 100 \frac{\psi(y) - \psi(y)_{Vreman}}{\psi(y)_{Vreman}} 
\end{eqnarray}
and
\begin{eqnarray}
||\epsilon_{\psi}|| &:=& \frac{1}{2h}\int_{-h}^{h} \!
                          \epsilon_{\psi} \: dy 
\end{eqnarray}
for each quantity $\psi$ = $U_{mean}$, $u_{rms}$, $v_{rms}$ and $w_{rms}$
in Table \ref{table: dns180comp}.
All statistics are within one percent of the results of \cite{vreman2014}.
The results indicate that increasing the number of Schwarz iterations 
and resolution leads to better comparison, as expected.

\begin{table}[h]
\begin{center}
 \begin{tabular}{|p{28mm} | c | c | c | c|}
 \hline
  & $U_{mean}$ & $u_{rms}$ & $v_{rms}$ & $w_{rms}$ \\
 \hline
 Monodomain, $N=9$ & 0.17  & 1.00 & 0.60  & 0.57 \\
 \hline
 Monodomain, $N=11$ & 0.16  & 0.46 & 0.49  & 0.71 \\
 \hline
 Overlapping, $N=9$, $\kappa_{iter}=1$, $m$ = 1 & 0.43  & 1.69 & 0.89  & 0.77 \\
 \hline
 Overlapping, $N=9$, $\kappa_{iter}=4$, $m$ = 3 & 0.21  & 0.92 & 0.60  & 1.05 \\
 \hline
 Overlapping, $N=11$, $\kappa_{iter}=1$, $m$ = 1 & 0.17  & 1.58 & 0.74  & 0.31 \\
 \hline
 Moser et. al. & 0.04  & 0.15 & 0.52  & 1.62 \\
 \hline
\end{tabular}
\vspace{-7mm}
\caption{Relative \% difference for channel flow results compared
with Vreman \& Kuerten \cite{vreman2014}. \vspace*{-.14in}}
\label{table: dns180comp}
\end{center}
\end{table}

\subsection{Heat Transfer Enhancement}
\begin{figure}[h] \begin{center} 
$\begin{array}{ccc}
\includegraphics[height=30mm]{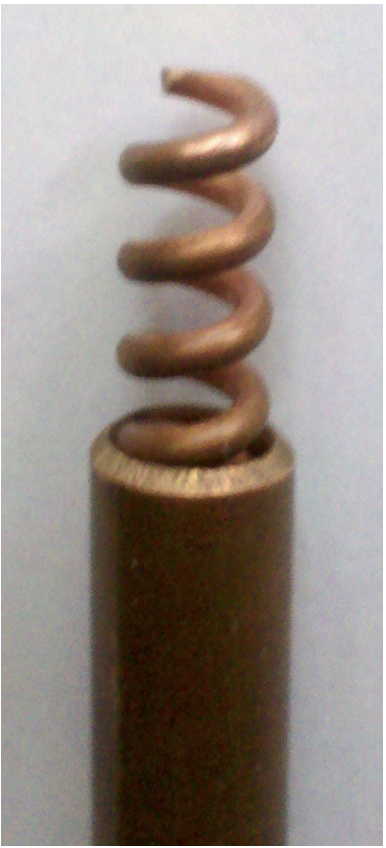} & \hspace{1em} 
\includegraphics[height=30mm]{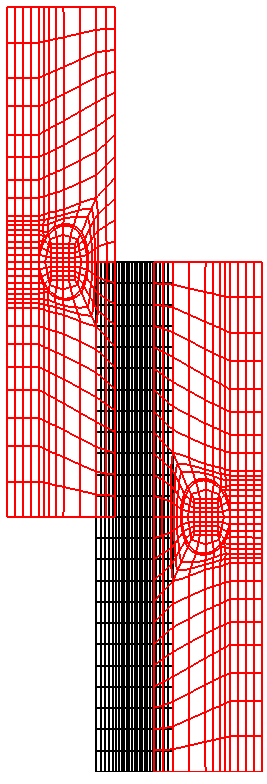} & \hspace{1em} 
\includegraphics[height=30mm]{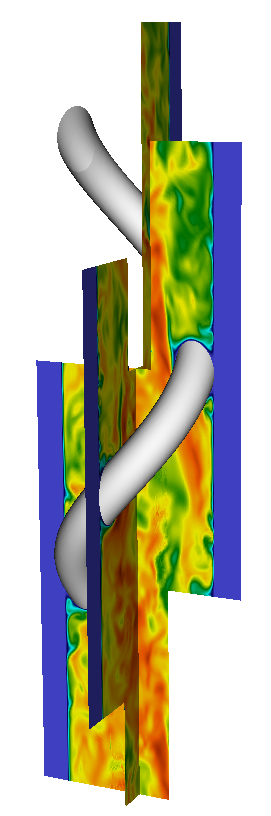} \\
\end{array}$ 
\end{center} 
\vspace{-7mm}
\caption{(left) Wire-coil insert \cite{collins2002}, (center) overlapping
spectral element meshes, and (right) velocity magnitude contours.}
\label{fig:helixmesh}
\end{figure}

The effectiveness of wire-coil inserts to increase heat transfer in pipe flow
has been studied through an extensive set of experiments by Collins {\em et
al.} at Argonne National Laboratory \cite{collins2002}.  Monodomain spectral
element simulations for this configuration based on 2D meshes that were
extruded and twisted were described in \cite{goering2016}.  Here, we consider
an overlapping mesh approach pictured in Fig. \ref{fig:helixmesh}, in which a
2D mesh is extruded azimuthally with a helical pitch.  The singularity at the
pipe centeriline is avoided by using a second (overlapping) mesh to model the
central flow channel.  The overlapping mesh avoids the topological constraint
of mesh conformity and leads to better mesh quality. Mesh smoothing
\cite{mittal2018} further improves the conditioning of the system for the
pressure Poisson equation. This approach also allows us to consider noncircular
(e.g., square) casings, which are not readily accessible with the monodomain
approach.  As a first step towards these more complex configurations, we
validate our Schwarz-SEM method with this real-world turbulent heat transfer
problem.  

Here, we present results from calculations for two different pitches of the
wire-coil insert to demonstrate the accuracy of our method.  The geometric
parameters were $e/p=0.0940$ (long-pitch) and $e/p=0.4273$ (optimal-pitch),
with $e/D=0.2507$, where $e$ and $p$ are the respective wire diameter and pitch
and $D$ is the inner diameter of the pipe.  The Reynolds number of the flow is
$UD/\nu=5300$ and the Prandtl number is 5.8. 
Figure \ref{fig:helixmesh} shows a typical wire-coil insert (left),
overlapping meshes used to discretize the problem (center), 
and a plot of velocity magnitude for flow at $Re=5300$ (right).

\begin{figure}[h] 
\begin{center}
$\begin{array}{cc}
\includegraphics[width=30mm]{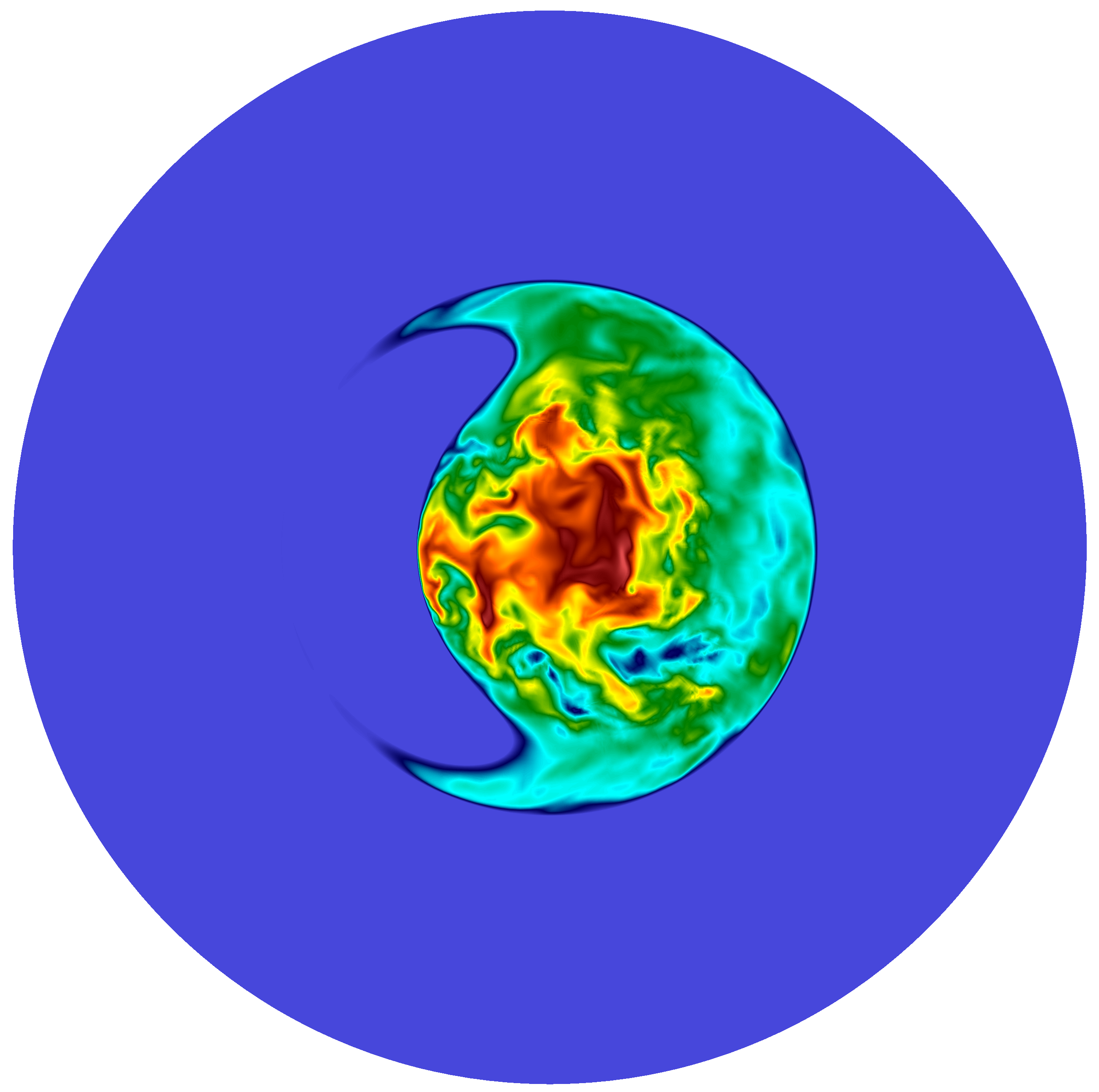} & \hspace{1em} 
\includegraphics[width=30mm]{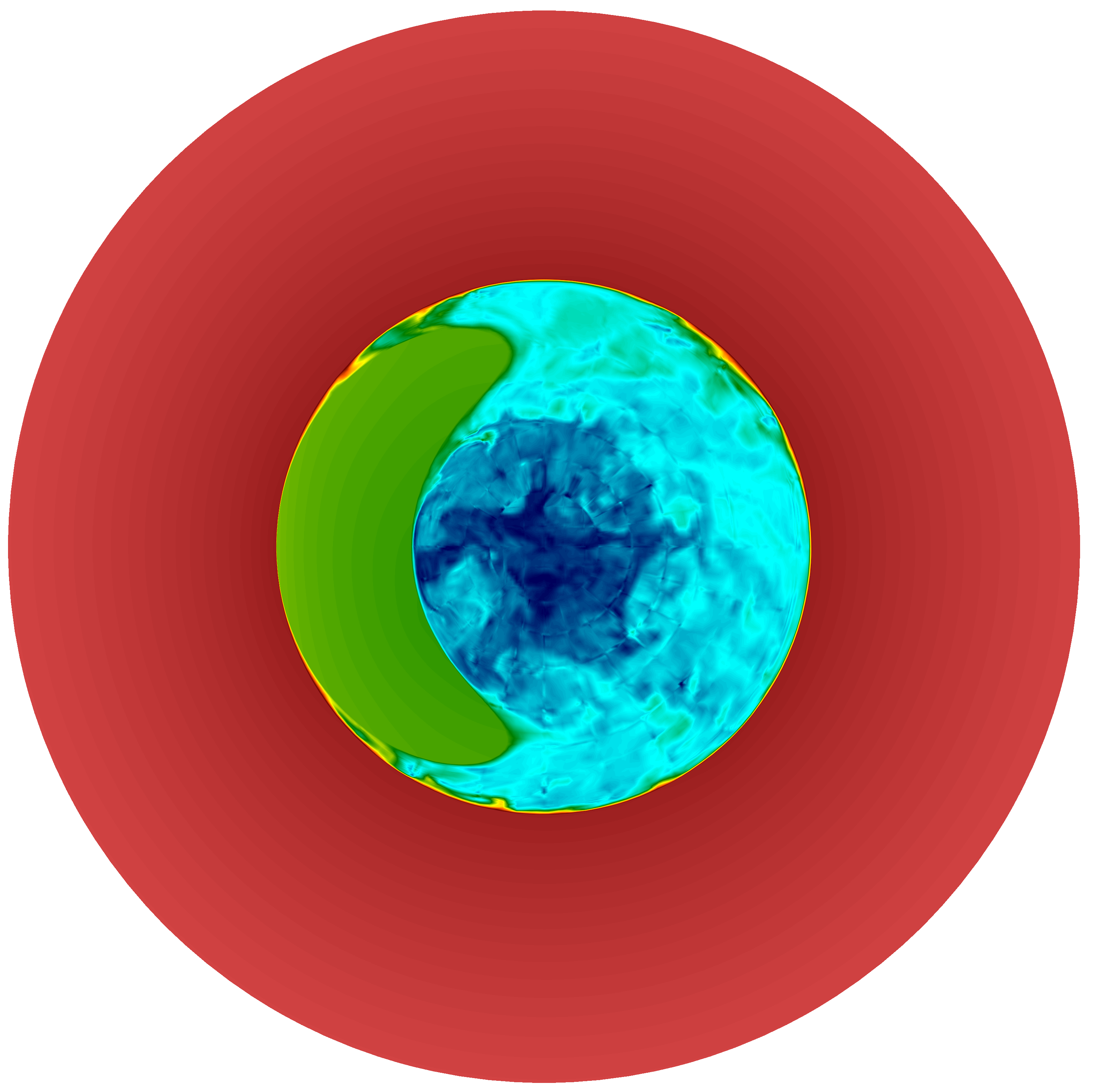} \\
\multicolumn{2}{c}{\includegraphics[width=60mm]{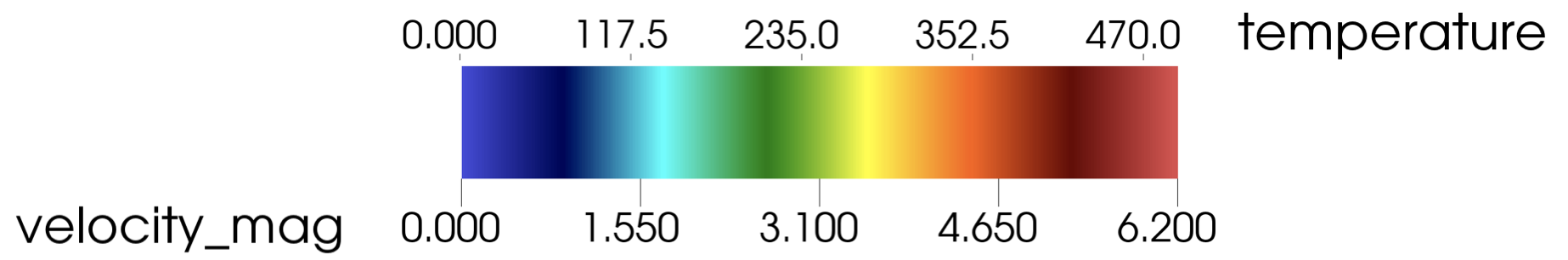}}
\end{array}$ 
\end{center} 
\vspace{-7mm}
\caption{Velocity magnitude (left) and temperature 
slice of the wire-coil insert of optimal pitch.} 
\label{fig:helixveltemp}
\end{figure}

For $e/p=0.0940$, the Nusselt number $Nu$ was determined to be 100.25 by
experimental calculation, 112.89 by the monodomain calculation, and 113.34 by
the overlapping-grid calculation. For $e/p=0.4273$, $Nu$ was determined to be
184.25 by the experiment, and 186 by the overlapping-grid calculation. The
overlapping grid calculations were spatially converged and used
$\kappa_{iter}=4$ and $m=3$. Figure \ref{fig:helixveltemp} shows a slice of the
velocity and temperature contours for the optimal pitch calculation.

\section{Conclusion}
We have presented a scalable Schwarz-SEM method for incompressible flow
simulation.  Use of an extended \emph{gslib-findpts} library for donor-element
identification and for data interpolation obviated the need for direct
development of MPI code, even in the presence of multiple overlapping meshes.
Strong scaling tests show that the parallel performance meets theoretical
expectations.  We have introduced mass-flux corrections to ensure
divergence-free flow, even in the presence of interpolation error.  We
demonstrated exponential convergence for the method and showed excellent
agreememnt in several complex three-dimensional flow configurations.

\section{Acknowledgments}
This work was supported by the U.S. Department of Energy,
Office of Science, the Office of Advanced Scientific Computing Research, under
Contract DE-AC02-06CH11357. An award of computer time on Blue Waters was
provided by the National Center for Supercomputing Applications. Blue Waters is
a sustained-petascale HPC and is a joint effort of the University of Illinois
at Urbana-Champaign and its National Center for Supercomputing Applications.
The Blue Waters sustained-petascale computing project is supported by the
National Science Foundation (awards OCI-0725070 and ACI-1238993) and the state
of Illinois. This research used resources of the Argonne Leadership Computing
Facility, which is a DOE Office of Science User Facility.



\section{References}
\tiny
\bibliographystyle{elsarticle-num} 
\bibliography{lit}

\begin{thebibliography}{10}
\expandafter\ifx\csname url\endcsname\relax
  \def\url#1{\texttt{#1}}\fi
\expandafter\ifx\csname urlprefix\endcsname\relax\def\urlprefix{URL }\fi
\expandafter\ifx\csname href\endcsname\relax
  \def\href#1#2{#2} \def\path#1{#1}\fi

\bibitem{dfm02}
M.~O. Deville, P.~F. Fischer, E.~H. Mund, High-order methods for incompressible
  fluid flow, Vol.~9, Cambridge University Press, 2002.

\bibitem{dutta2016}
S.~Dutta, P.~Fischer, M.~H. Garcia, {Large Eddy Simulation (LES) of flow and
  bedload transport at an idealized 90-degree diversion: Insight into Bulle
  effect}, River Flow 2016: Iowa City, USA, July 11-14, 2016 (2016) 101.

\bibitem{hosseini2016}
S.~M. Hosseini, R.~Vinuesa, P.~Schlatter, A.~Hanifi, D.~S. Henningson, Direct
  numerical simulation of the flow around a wing section at moderate reynolds
  number, International Journal of Heat and Fluid Flow 61 (2016) 117--128.

\bibitem{merzari2017}
E.~Merzari, A.~Obabko, P.~Fischer, Spectral element methods for liquid metal
  reactors applications, arXiv preprint arXiv:1711.09307.

\bibitem{sao80}
S.~A. Orszag, Spectral methods for problems in complex geometrics, in:
  Numerical methods for partial differential equations, Elsevier, 1979, pp.
  273--305.

\bibitem{patera84}
A.~T. Patera, A spectral element method for fluid dynamics: laminar flow in a
  channel expansion, Journal of computational Physics 54~(3) (1984) 468--488.

\bibitem{Rogersbestpractices}
W.~Chan, R.~Gomez, Rogers se, buning pg. best practices in overset grid
  generation. aiaa\# 2002-3191, in: 32nd Fluid Dynamics Conference, St. Louis
  MI, 2002.

\bibitem{steger1983}
J.~L. Steger, F.~C. Dougherty, J.~A. Benek, A chimera grid scheme.[multiple
  overset body-conforming mesh system for finite difference adaptation to
  complex aircraft configurations].

\bibitem{angel2018}
J.~B. Angel, J.~W. Banks, W.~D. Henshaw, A high-order accurate fdtd scheme for
  maxwell's equations on overset grids, in: Applied Computational
  Electromagnetics Society Symposium (ACES), 2018 International, IEEE, 2018,
  pp. 1--2.

\bibitem{chandar2018comparative}
D.~D. Chandar, B.~Boppana, V.~Kumar, A comparative study of different overset
  grid solvers between openfoam, starccm+ and ansys-fluent, in: 2018 AIAA
  Aerospace Sciences Meeting, 2018, p. 1564.

\bibitem{koblitz2017}
A.~Koblitz, S.~Lovett, N.~Nikiforakis, W.~Henshaw, Direct numerical simulation
  of particulate flows with an overset grid method, Journal of Computational
  Physics 343 (2017) 414--431.

\bibitem{cd2012v7}
S.-C. CD-adapco, V7. 02.008, User Manual.

\bibitem{nicholsoverflowmanual}
R.~H. Nichols, P.~G. Buning, User’s manual for overflow 2.1, University of
  Alabama and NASA Langley Research Center.

\bibitem{overturemanual}
F.~Bassetti, D.~Brown, K.~Davis, W.~Henshaw, D.~Quinlan, Overture: an
  object-oriented framework for high performance scientific computing, in:
  Proceedings of the 1998 ACM/IEEE conference on Supercomputing, IEEE Computer
  Society, 1998, pp. 1--9.

\bibitem{coder2017contributions}
J.~G. Coder, D.~Hue, G.~Kenway, T.~H. Pulliam, A.~J. Sclafani, L.~Serrano,
  J.~C. Vassberg, Contributions to the sixth drag prediction workshop using
  structured, overset grid methods, Journal of Aircraft (2017) 1--14.

\bibitem{henshaw1994}
W.~D. Henshaw, A fourth-order accurate method for the incompressible
  navier-stokes equations on overlapping grids, Journal of computational
  physics 113~(1) (1994) 13--25.

\bibitem{brazell2016overset}
M.~J. Brazell, J.~Sitaraman, D.~J. Mavriplis, An overset mesh approach for 3d
  mixed element high-order discretizations, Journal of Computational Physics
  322 (2016) 33--51.

\bibitem{crabill2016}
J.~A. Crabill, J.~Sitaraman, A.~Jameson, A high-order overset method on moving
  and deforming grids, in: AIAA Modeling and Simulation Technologies
  Conference, 2016, p. 3225.

\bibitem{aarnes2018high}
J.~Aarnes, N.~Haugen, H.~Andersson, High-order overset grid method for
  detecting particle impaction on a cylinder in a cross flow, arXiv preprint
  arXiv:1805.10039.

\bibitem{cambier2013onera}
L.~Cambier, S.~Heib, S.~Plot, The onera elsa cfd software: input from research
  and feedback from industry, Mechanics \& Industry 14~(3) (2013) 159--174.

\bibitem{merrill2016}
B.~E. Merrill, Y.~T. Peet, P.~F. Fischer, J.~W. Lottes, A spectrally accurate
  method for overlapping grid solution of incompressible navier--stokes
  equations, Journal of Computational Physics 307 (2016) 60--93.

\bibitem{malm13}
J.~Malm, P.~Schlatter, P.~F. Fischer, D.~S. Henningson, Stabilization of the
  spectral element method in convection dominated flows by recovery of
  skew-symmetry, Journal of Scientific Computing 57~(2) (2013) 254--277.

\bibitem{fischer17}
P.~Fischer, M.~Schmitt, A.~Tomboulides, Recent developments in spectral element
  simulations of moving-domain problems, in: Recent Progress and Modern
  Challenges in Applied Mathematics, Modeling and Computational Science,
  Springer, 2017, pp. 213--244.

\bibitem{peet2012}
Y.~T. Peet, P.~F. Fischer, Stability analysis of interface temporal
  discretization in grid overlapping methods, SIAM Journal on Numerical
  Analysis 50~(6) (2012) 3375--3401.

\bibitem{fox88}
G.~Fox, W.~Furmanski, Hypercube algorithms for neural network simulation: the
  crystal\_accumulator and the crystal router, in: Proceedings of the third
  conference on Hypercube concurrent computers and applications: Architecture,
  software, computer systems, and general issues-Volume 1, ACM, 1988, pp.
  714--724.

\bibitem{findpts}
A.~Noorani, A.~Peplinski, P.~Schlatter, Informal introduction to program
  structure of spectral interpolation in nek5000.

\bibitem{fischer2015scaling}
P.~F. Fischer, Scaling limits for pde-based simulation, in: 22nd AIAA
  Computational Fluid Dynamics Conference, 2015, p. 3049.

\bibitem{escudier1984}
M.~Escudier, Observations of the flow produced in a cylindrical container by a
  rotating endwall, Experiments in fluids 2~(4) (1984) 189--196.

\bibitem{sotiropoulos1998}
F.~Sotiropoulos, Y.~Ventikos, Transition from bubble-type vortex breakdown to
  columnar vortex in a confined swirling flow, International journal of heat
  and fluid flow 19~(5) (1998) 446--458.

\bibitem{moser1999}
R.~D. Moser, J.~Kim, N.~N. Mansour, Direct numerical simulation of turbulent
  channel flow up to re $\tau$= 590, Physics of fluids 11~(4) (1999) 943--945.

\bibitem{vreman2014}
A.~Vreman, J.~G. Kuerten, Comparison of direct numerical simulation databases
  of turbulent channel flow at re $\tau$= 180, Physics of Fluids 26~(1) (2014)
  015102.

\bibitem{collins2002}
J.~T. Collins, C.~M. Conley, J.~N. Attig, M.~M. Baehl, Enhanced heat transfer
  using wire-coil inserts for high-heat-load applications., Tech. rep., Argonne
  National Lab., IL (US) (2002).

\bibitem{goering2016}
A.~Y. Goering, Numerical investigation of wire coil heat transfer augmentation,
  Ph.D. thesis (2016).

\bibitem{mittal2018}
K.~Mittal, P.~Fischer, Mesh smoothing for the spectral element method, Journal
  of Scientific Computing (2018) 1--22.

\end{thebibliography}

\end{document}